\documentclass[journal=jasms,manuscript=article]{achemso}

\usepackage{epsf, latexsym, color, epsfig}
\usepackage{amssymb, amsmath}

\newcommand{\ket}[1]{| #1 \rangle}

\newcommand{\ra}{{\rightarrow}}

\newcommand{\be}{\begin{equation}}
\newcommand{\ee}{\end{equation}}
\newcommand{\ba}{\begin{eqnarray}}
\newcommand{\ea}{\end{eqnarray}}

\newcommand\prl{Phys.~Rev.~Lett.~}
\newcommand\pra{Phys.~Rev.~A~}
\newcommand\prb{Phys.~Rev.~B~}
\newcommand\rmp{Rev.~Mod.~Phys.~}

\title{Interferometric mass spectrometry}

\author{Radu Ionicioiu}
\affiliation{Horia Hulubei National Institute of Physics and Nuclear Engineering, 077125 Bucharest--M\u agurele, Romania}
\email{r.ionicioiu@theory.nipne.ro}

\begin{document}
%%%%%%%%%%%%%%%%%%%%%%%%%%%%%%%%%%%%%%%%%%%%%%%%%%%%%%%%%%%%%%%%%%%%%
\begin{tocentry}

\includegraphics[width= 7.8cm]{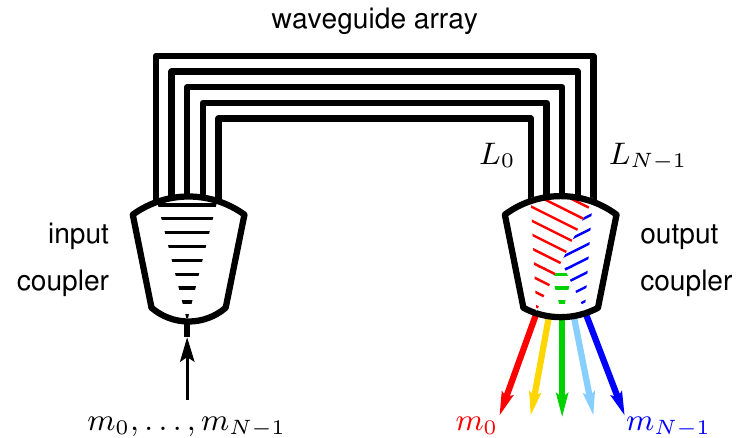}

~\\
{\bf Title:} Interferometric mass spectrometry

{\bf Author:} Radu Ionicioiu

{\bf Synopsis:} Interferometric mass spectrometry (Interf-MS). The input coupler prepares the wavefunction into an equal superposition of $N$ paths. Subsequently, the wave-function propagates through a multi-path interferometer. The paths have different lengths and introduce distinct phase shifts for different mass species. Due to constructive interference at the output coupler, each mass species exits from its own output, i.e., $m_i$ exits from output $i$.

\end{tocentry}
%%%%%%%%%%%%%%%%%%%%%%%%%%%%%%%%%%%%%%%%%%%%%%%%%%%%%%%%%%%%%%%%%%%%%

\begin{abstract}
Accelerator mass spectrometry (AMS) is a widely-used technique with multiple applications, including geology, molecular biology and archeology. In order to achieve a high dynamic range, AMS requires tandem accelerators and large magnets, which thus confines it to big laboratories. Here we propose interferometric mass spectrometry (Interf-MS), a novel method of mass separation which uses quantum interference. Interf-MS employs the wave-like properties of the samples, and as such is complementary to AMS, in which samples are particle-like. This complementarity has two significant consequences: (i) in Interf-MS separation is performed according to the absolute mass $m$, and not to the mass-to-charge ratio $m/q$, as in AMS; (ii) in Interf-MS the samples are in the low-velocity regime, in contrast to the high-velocity regime used in AMS. Potential applications of Interf-MS are compact devices for mobile applications, sensitive molecules that break at the acceleration stage and neutral samples which are difficult to ionise. 
\end{abstract}

\section{Introduction}

Mass spectrometry is a widespread tool with applications in several fields like chemistry, molecular biology, geology and archeology \cite{muller}. A well-known example is carbon dating, a technique used to determine the age of organic samples by measuring the relative abundances of $^{12}$C and $^{14}$C isotopes. Due to its precision and conceptual simplicity, carbon dating had a huge impact and revolutionised several fields like archeology, palaeontology and anthropology.

One of most precise techniques is accelerator mass spectrometry (AMS). In AMS the samples (i.e., the mass species we want to separate) are first ionised, then accelerated to high velocities using a tandem accelerator. Electric and magnetic fields are used in velocity filters and in mass analysers to separate different mass species according to their mass-to-charge ratios $m/q$. After the separation stage, an array of particle detectors count the number of atoms for each species.

Since samples are, after all, quantum systems, one question arises. Can we use the wave-like properties for mass separation? Here we answer this question in the affirmative. Quantum superposition and matter-wave interferometry have been experimentally demonstrated for electrons \cite{electron, electron2}, neutrons \cite{neutron}, atoms \cite{atom} and increasingly large molecules \cite{molecules}, including C60 \cite{c60}, molecules \cite{kapitza, bragg, 25kda}, molecular clusters \cite{waals, clusters} and polypeptides \cite{polypeptide}.

In this article we describe a novel mass spectrometry method which uses quantum interference as a mass-separation mechanism. Interferometric sorting has been used previously for spin separation \cite{spin}, sorting quantum systems \cite{sorter}, orbital angular momentum of photons \cite{sorter_OAM, berk, mirh} and radial modes of light \cite{radial_s1, radial_s2}. We start with a brief overview of the the standard method, namely accelerator mass spectrometry. We then introduce interferometric sorting, first for two mass species and then for the general case of $N$ species.

\section{Methods}

\subsection{Accelerator mass spectrometry}

\begin{figure}[t]
  \includegraphics[width= 5cm]{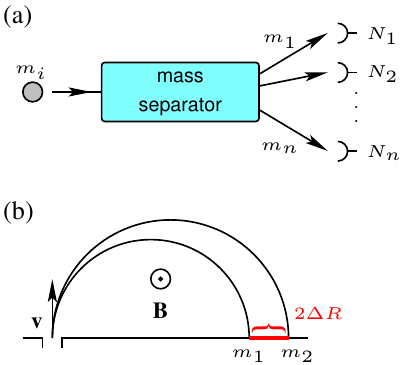}
  \caption{(a) Schematics of a mass spectrometer. A mass analyser (separator) separates different mass species into distinct spatial channels; each channel has a particle counter measuring the number of particles $N_i$. (b) In AMS the analyser uses magnetic fields for mass separation.}
  \label{MS}
\end{figure}

The basic idea of a mass spectrometer is shown in Fig.\ref{MS}. The samples of masses $m_i$ enter a mass separator which directs each mass species (isobars) into distinct spatial channels. Each channel $i$ has a particle counter which registers the number of particles $N_i$ for a given period of time. The mass spectrum is calculated from the relative counting frequencies $N_1, N_2, \ldots$, where $N_i$ is the number of particles in detector $i$, corresponding to mass $m_i$. The detectors themselves are mass-insensitive, i.e., they do not distinguish between different mass species. Thus the key element is the mass separator which sorts the particles into distinct spatial channels (paths) according to their masses.

In AMS the mass analyser (or separator) uses the Lorentz force acting on charged particles. From $q vB= mv^2/R$, we have
\be
R= \frac{mv}{qB} 
\ee
A velocity filter placed before the analyser ensures that all samples have the same velocity $v$. Consider two mass species $(m_1, q_1)$ and $(m_2, q_2)$. Then the separation between the two species is $2\Delta R$, with
\be
\Delta R= \frac{v}{B} \left( \frac{m_2}{q_2}- \frac{m_1}{q_1} \right)
\ee
There are two important points to note here. First, the Lorentz force separates species according to the mass-to-charge ratio $m/q$, and not to the mass $m$. Therefore the samples have to be electrically charged, as the Lorentz force cannot separate neutral particles. Second, for a given magnetic field $B$ the separation between two samples is proportional to the velocity $v$. AMS uses TANDEM accelerators to achieve a very high dynamical range in order to detect minute amounts of isotopes (like $^{14}$C) against the background.

\subsection{Interferometric sorting: two mass species}

Interferometric sorting uses wave-like properties of quantum systems (atoms, molecules, proteins etc) to separate the samples according to their mass. An interferometric mass separator uses constructive/destructive interference to separate different species $m_i$ into distinct spatial channels (paths). First we discuss the simpler case of sorting two mass species, and then we generalise to $N$ species.

\begin{figure}[t]
  \includegraphics[width= 6cm]{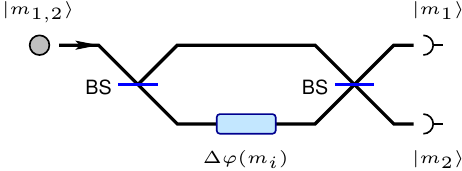}
  \caption{Interferometric mass spectrometry (Interf-MS) for two mass species. A Mach-Zehnder interferometer has a mass-dependent phase-shift $\Delta\varphi(m_i):= \Delta\varphi_i$ in one arm, such that the mass species $m_{1,2}$ acquire different phases $\Delta\varphi_{1,2}$. At the second beam-splitter $m_1$ interferes constructively in output 1 and destructively in output 2, whereas $m_2$ interferes constructively in output 2 and destructively in output 1.}
  \label{IMS2}
\end{figure}

Let us consider the problem of sorting two mass species $m_1, m_2$. The general scheme is shown in Figure \ref{IMS2}. It consists of a Mach-Zehnder interferometer (MZI) with a mass-dependent phase shift in the lower arm:
\be
\ket{m_i} \ra\ e^{i\Delta\varphi_i}\ket{m_i}, \ \ i=1,2
\ee
To achieve (perfect) interferometric separation of the two mass species, we choose the phase shifts such that $m_1$ has constructive interference in output 1, whereas $m_2$ has constructive interference in output 2. This implies the phase-shift induces a phase $\Delta\varphi_1= 0$ for $m_1$ and $\Delta\varphi_2= \pi$ for $m_2$:
\ba
\nonumber
\ket{m_1} &\ra& \ket{m_1} \\
\ket{m_2} &\ra& -\ket{m_2} 
\ea

In quantum information parlance, the mass-dependent phase-shift acts like a controlled-$Z$ gate between the mass and the path degrees of freedom \cite{sorter}. Therefore the interferometer acts as a CNOT gate between mass and path degrees of freedom. This insight is important and will provide the solution to the general case of sorting $N$ species. Next we discuss how to implement the mass-dependent phase shift.

\noindent{\bf Mass-dependent phase shift}. Consider a quantum system (atom, molecule etc), incident on a Mach-Zehnder interferometer, Fig.\ref{IMS2}; we will use ``particle'' as a short-hand for quantum system. After the first beamsplitter, the particle is in superposition of being in the two paths. We assume the paths have lengths $L_1$ and $L_2$, respectively. The particle interferes with itself at the second beamsplitter. The phase difference between the two paths is
\be
\Delta\varphi= \frac{2\pi \Delta L}{\lambda}= \frac{2\pi \Delta L}{h} mv
\ee
with $\Delta L= L_2 - L_1$; $\lambda= h/mv$ is the de Broglie wavelength of the particle of mass $m$ and velocity $v$; $h$ is Planck's constant.

Now suppose we have two mass species $m_{1,2}$ entering the interferometer with velocities $v_{1,2}$, respectively. Then
\be
\frac{\Delta \varphi_1}{\Delta \varphi_2}= \frac{m_1 v_1}{m_2 v_2} 
\ee

To achieve {\em perfect interferometric sorting}, we design the interferometer such that $m_1$ ($m_2$) exits on output 1 (output 2) with unit probability. This implies the following relations:
\begin{subequations}
\ba
\label{phi1}
\Delta\varphi_1 &=& 2 k_1 \pi \\
\label{phi2}
\Delta\varphi_2 &=& (2k_2+ 1)\pi
\ea
\end{subequations}
with arbitrary $k_1, k_2\in {\mathbb Z}$. Thus the conditions for interferometric sorting are (not all equations are independent):
\begin{subequations}
\label{sorting}
\ba
\frac{m_1 v_1}{m_2 v_2} &=& \frac{2k_1}{2k_2+1} \\
\Delta L &=& k_1 \lambda_1= k_1 \frac{h}{m_1 v_1} \\
\Delta L &=& (k_2+ \tfrac{1}{2}) \lambda_2= (k_2+ \tfrac{1}{2}) \frac{h}{m_2 v_2}
\ea
\end{subequations}
We assume there is a velocity filter before the interferometer, hence $v_1= v_2= v$ and the first equation becomes $m_1/m_2= 2k_1/(2k_2+1)$. As a velocity filter we can use a Wien filter (for charged species) or a pair of mechanichal choppers (in the low velocity regime, for non-charged species).

Clearly, the parameter we need to control accurately is the path difference $\Delta L$ between the two arms. We can see this in the the next example.

\noindent{\bf Example.} Carbon dating is a standard method to determine the age of organic materials like wood, bone etc. In carbon dating we measure the relative abundances of two carbon isotopes, $^{12}$C and $^{14}$C. We have $m_1= 1.99\times 10^{-26}$\,kg ($^{12}$C) and $m_1/m_2= 6/7$. This gives $k_1=k_2= 3$ and $\Delta L= 3 \lambda_1$. For $v= v_{1,2}= 100$ m/s we obtain $\Delta L\approx$1\,nm. If we slow down the samples to $v=1$ m/s, then $\Delta L\approx 0.1\,\mu$m.

The challenge is to accurately control (and stabilise) the path difference $\Delta L$ between the two arms of the MZI. Commercially available piezo-stages used for nano-positioning (e.g., in AFM) have sub-nanometer resolution $\sim$0.3\,nm \cite{piezo}. For the previous example ($^{12}$C, $v=1$ m/s), a path difference $\Delta L= 0.3$\,nm corresponds to a phase difference $\Delta\varphi= 1.88 \times 10^{-2}$ rad.

\subsection{Interferometric sorting: $N$ mass species}

We now turn to the general case. The problem of sorting quantum systems according to an arbitrary (discrete) degree of freedom has been studied in \cite{sorter}. Here we apply similar considerations for sorting $N$ mass species $m_0, \ldots, m_{N-1}$.

\begin{figure}[t]
  \includegraphics[width= 7cm]{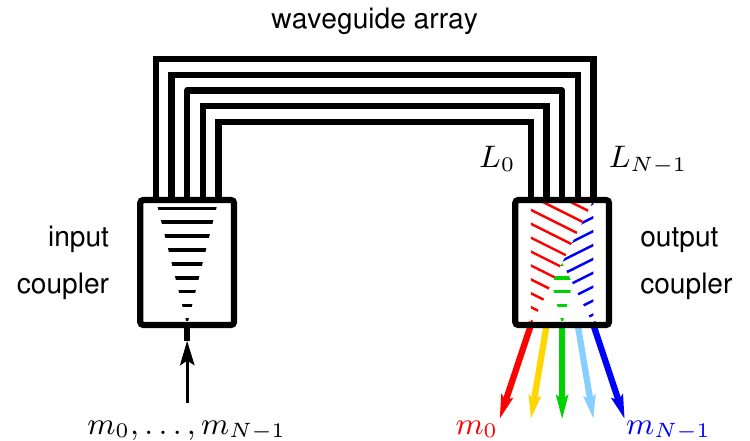}
  \caption{Interferometric mass separation for $N$ species. The input coupler puts the wavefunction into an equal superposition of $N$ paths. The wave-function then propagates through a multi-path interferometer. The paths have different lengths and introduce distinct phase shifts for different mass species. Due to constructive interference at the output coupler, each mass species exits from its own output, i.e., $m_i$ exits from output $i$.}
  \label{IMS_n}
\end{figure}

The device for interferometric sorting is a multi-path Mach-Zehnder interferometer and consists of an input coupler, a waveguide array and an output coupler, Fig.~\ref{IMS_n}. All mass species enter through the same port of the input coupler. The waveguide array has $N$ waveguides of lengths $L_0, \ldots, L_{N-1}$. The output coupler has $N$ output waveguides used for sorting the $N$ species. The input (output) coupler is a free-propagating region and implements the discrete Fourier transform on spatial modes (paths) \cite{lopez, bachmann, zhou}. The input coupler ensures the quantum systems enter the waveguide array in an equal superposition of paths, $\tfrac{1}{\sqrt{N}} \sum_{i=0}^{N-1} \ket{i}$. At the output coupler all the waves coming from the waveguide arrays interfere with different phases.

For sorting, we require that mass $m_k$ exits from output port $k$ with unit probability, i.e., there is constructive interference in port $k$ and destructive interference in all other ports $j\ne k$. We assume that all species have the same velocity $v$, hence there is a velocity filter before the analyser. A species $m_k$ propagating along waveguide $L_s$ acquires a phase shift
\be
\varphi_{k, s}= 2\pi \frac{L_s}{\lambda_k}
\ee
with $\lambda_k= h/m_k v$ its de Broglie wavelength. Since species $m_k$ exits on port $k$ with probability $p=1$, this requires the phase differences between each arm of the interferometer and the reference 0-th arm to satisfy the conditions \cite{sorter}:
\be
\Delta \varphi_{k, s}:= \varphi_{k, s}- \varphi_{k, 0}= \frac{2\pi}{N} ks + 2\pi n_{k,s}
\label{delta_phi}
\ee
for all $k, s= 0, \ldots, N-1$ and arbitrary integers $n_{k,s} \in \mathbb{Z}$, $n_{k,0}= 0$. These are the conditions for {\it perfect interferometric sorting} which generalise eqs.~\eqref{phi1}, \eqref{phi2}. We have
\begin{eqnarray}
\Delta L_s:= L_s- L_0&=& \lambda_k \left[ \frac{ks}{N} + n_{k,s} \right] \\
\label{deltaLs}
&=& \frac{h}{m_k v} \left[ \frac{ks}{N} + n_{k,s} \right]
\end{eqnarray}
for all $k, s= 0, \ldots, N-1$ and arbitrary $n_{k,s} \in \mathbb{Z}$, $n_{k,0}= 0$.

Experimentally we need to control the path differences $\Delta L_s$ of the waveguide array in order to induce the appropriate phase-shifts. Since $\Delta L_s \sim v^{-1}$, this implies that we need to work in the low velocity regime. For the $^{12}$C example, $m_0= 1.99\times 10^{-26}$ and taking $\Delta L_1= 1$\, nm, we have $v= n_{0,1} h/m_0 \Delta L_1$; modulo an integer $n_{0,1}$, $v= 33.3$ m/s, consistent with the low-velocity regime.

From eq.~\eqref{delta_phi} we see that, given a mass species $m_k$, all the phases $\Delta \varphi_{k, s}$, $s=0,\ldots, N-1$, are different (modulo $2\pi$) iff $(k, N)$ are coprime. Consequently, if $N$ is a prime number, all phases $\Delta \varphi_{k, s}$ are distinct (mod $2\pi$), for any mass species $m_k$, $k=0,\ldots, N-1$.

\subsection{Discrete Fourier Transform}

A key element of interferometric sorting is the input (output) coupler implementing the discrete Fourier transform (DFT) $F$ on paths (spatial modes) $\ket{k}$:
\be
F: \ket{k} \rightarrow \frac{1}{\sqrt N} \sum_{j=0}^{N-1} \omega^{kj} \ket{j}
\label{dft}
\ee
with $\omega= e^{2\pi i/N}$ a root of unity of order $N$. Thus the output of the DFT $F$, eq.~\eqref{dft}, is a wave in a supperposition of all paths with the same probability amplitude $\frac{1}{\sqrt N}$ (hence probability $1/N$), but with different phases given by $\omega^{kj}$. For the $k=0$ input, all the phases are 1 and the action of $F$ is simply $\ket{0} \rightarrow \frac{1}{\sqrt N} \sum_{j=0}^{N-1} \ket{j}$, i.e., a wave in equal superposition of outputs and with same phases.

For the simplest case $N=2$, the Fourier transform is given by the Hadamard matrix $H= \frac{1}{\sqrt 2}\begin{bmatrix} 1 & 1 \\ 1 & -1 \end{bmatrix}$ and in optics is implemented by a 50/50 beam-splitter.

\begin{figure}[t]
  \includegraphics[width= 7cm]{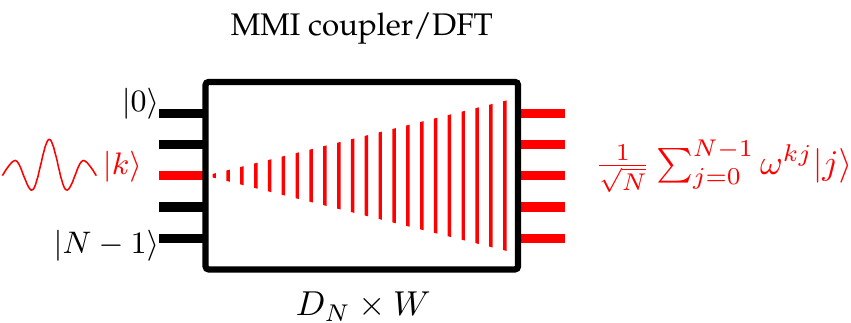}
  \caption{The input (output) coupler is equivalent to the discrete Fourier Transform on $N$ paths (spatial modes). The coupler is implemented by a multi-mode interference (MMI) device of length $D_N$ and width $W$.}
  \label{MMI}
\end{figure}

In order to implement the DFT for the general case of Interf-MS, we take our inspiration from waveguide optics, as both phenomena use interference of quantum waves. In waveguide optics a multimode interference (MMI) coupler implements the DFT \cite{lopez, bachmann, zhou}. The MMI consists of a rectangular, free-propagation region of width $W$ and length $D_N$, with $N$-inputs and $N$-outputs, as in Fig.~\ref{MMI}. The device uses self-imaging inside the multimode waveguide. For a device with $N$ input waveguides and width $W$, the length required to implement the DFT (up to a permutation of the waveguides) is given by \cite{lopez, bachmann}
\be
D_N= \frac{4 W^2}{\lambda N}
\label{D_N}
\ee
where $\lambda= h/mv$ is the wavelength in vacuum. For $^{12}$C, $m= 1.99\times 10^{-26}$\,kg and $v= 1$ m/s, we take $W= 1\, \mu$m and $N=5$, the MMI (coupler) length is $D_5= 24\, \mu$m.

The MMI coupler is assumed to be planar. This implies that the thickness of the MMI should be similar to the thickness of the incoming/outgoing waveguides.

Theoretically, Interf-MS works for both charged and neutral particles. For charged particles, however, interference could be destroyed by mirror charges. In order to avoid this, the coupler should be dielectrically coated or made of a single block of dielectric, for example silicon.

\subsection{Errors}

So far we have discussed the ideal case. We now briefly analyse the effect of errors around the ideal parameters. One source of errors are fluctuations in length of the waveguide array. We assume that path $s$ of the interferometer has length $L'_s= L_s + \delta L_s$, where $\delta L_s$ is the fluctuation around the theoretical (ideal) value $L_s$. Thus species $m_k$ propagating along path $s$ acquires a phase $\varphi'_{k,s}= \varphi_{k,s} + \delta \varphi_{k,s}$, with $\varphi_{k,s}= 2\pi L_s /\lambda_k$ and $\delta \varphi_{k,s}= 2\pi \delta L_s /\lambda_k$. For the mass species $m_k$, the phase difference between path $s$ and path 0 is
\be
\Delta \varphi'_{k, s}:= \varphi'_{k, s}- \varphi'_{k, 0}= \Delta \varphi_{k, s} + \frac{2\pi}{\lambda_k} (\delta L_s - \delta L_0)
\ee
Consequently, if all the paths fluctuate by the same amount, $\delta L_s = \delta L_0, \forall s=0,\ldots, N-1$, then the perfect sorting conditions \eqref{delta_phi} are unchanged. This shows the robustness under global fluctuations, where all paths are modified equally.

Consider now uncorrelated errors. For an $N$-path interferometer, the phases in eq.~\eqref{delta_phi} are multiples of $\delta \varphi= \frac{2\pi}{N}$, which corresponds to a difference in path length
\be
\delta L= \frac{\lambda}{N}
\ee
As a rule of thumb, for $N$ species we need to control the path differences in the interferometer to order ${\cal O}(\lambda/N)$, with $\lambda= \min_k \lambda_k$ the smallest wavelength of the set. One way around this is to concatenate a series of separation stages, where each stage separates only between a small number of species.

\subsection{Signal leakage}

An important issue is to estimate the signal leakage from one channel (path) to another. In order to do this we will use a quantum information model of the device. As discussed before, the ideal mass sorter is a quantum $C(X_N)$ gate (controlled-$X_N$ gate), between the mass degree of freedom and the path degree of freedom. In quantum information both degrees of freedom are modelled as qudits, i.e., $N$-dimensional quantum systems. In our case we have the masses $m_k$, with the corresponding qudit state $\ket{k}$,  and the paths $\ket{s}$, with $k, s = 0, \ldots, N-1$. The action of the ideal sorter is
\be
C(X_N): \ \ket{k}_{mass} \ket{s}_{path} \mapsto \ket{k}_{mass} \ket{s\oplus k}_{path}
\ee
with $\oplus$ addition modulo $N$. For sorting, all mass species enter the device on the same path $k=0$, hence the action of the ideal sorter is
\be
C(X_N): \ \ket{k}_{mass} \ket{0}_{path} \mapsto \ket{k}_{mass} \ket{k}_{path}
\label{CX_ideal}
\ee
i.e., a species $m_k$ exits on the output (path) $k$ with probability 1.

\begin{figure}[t]
  \includegraphics[width= 16cm]{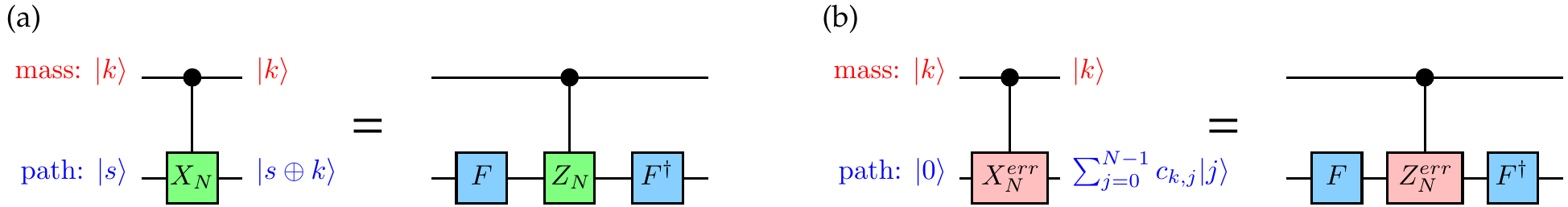}
  \caption{(a) An ideal sorter is equivalent to $C(X_N)= (I\otimes F^\dag) C(Z_N) (I\otimes F)$; $F, F^\dag$ act only on the path qudit. (b) Error model for the imperfect sorter $C(X_N^{err})= (I\otimes F^\dag) C(Z_N^{err}) (I\otimes F)$.}
  \label{Xd}
\end{figure}

We use the gate identity $C(X_N)= (I\otimes F^\dag) C(Z_N) (I\otimes F)$; here the Fourier gates $F, F^\dag$ act only on the path qudit, $I$ is the identity gate (on the first qudit) and $C(Z_N)$ is a controlled-$Z_N$ gate between mass and path qudits, see Fig.~\ref{Xd}(a).

We consider a simple error model in which errors result from path fluctuations $\delta L_s$ inside the $N$-path interferometer. The path errors generate phase errors, which are modelled by the $C(Z_N^{err})$ quantum gate. The ideal $C(Z_N)$ gate is
\be
C(Z_N): \ \ket{k}_{mass} \ket{s}_{path} \mapsto \omega^{sk}\, \ket{k}_{mass} \ket{s}_{path}
\ee
with $\omega= e^{2\pi i/N}$ a root of unity of order $N$ and $k, s= 0, \ldots, N-1$. The errors induce extra phases in the $C(Z_N^{err})$ gate:
\be
C(Z_N^{err}): \ \ket{k}_{mass} \ket{s}_{path} \mapsto e^{i\delta \varphi_{k,s}}\, \omega^{sk}\, \ket{k}_{mass} \ket{s}_{path}
\ee
where $e^{i\delta \varphi_{k,s}}$ are the phase errors induced by the path errors $\delta L_s$. One can prove that
\be
\delta \varphi_{k,s}= \frac{m_k}{m_0}\, \delta \varphi_{0,s}
\ee
and moreover $\delta \varphi_{k,0}= 0$; this follows from the fact that only relative phases with respect to the 0-path are relevant. Thus there are only $N-1$ independent phase errors $\delta \varphi_{0, s}$ corresponding to $N-1$ independent relative-path errors $\delta L_s, s= 1,\ldots, N-1$.

Suppose a mass species $m_k$ enters the imperfect interferometer. In contrast to the ideal case \eqref{CX_ideal}, now $m_k$ will not exit on path $k$ with probability 1; instead it will be in a quantum superposition of paths $s$, see Fig.~\ref{Xd}(b)
\be
C(X_N^{err}): \ \ket{k}_{mass} \ket{0}_{path} \mapsto \ket{k}_{mass} \sum_{s=0}^{N-1} c_{k, s} \ket{s}_{path}
\label{CX_error}
\ee
where $c_{k, s}$ are complex probability amplitudes. The probability of mass $m_k$ exiting from path $s$ is $p_{k,s}= |c_{k, s}|^2$ and this quantifies the leakage between channels (paths). In the ideal (no errors) case we have $c_{k,s}= p_{k,s}= \delta_{ks}$, the Kronecker delta function.

\noindent{\bf Example.} We now calculate analytically the probability amplitudes for $N=3$ mass species $m_0, m_1, m_2$. In this case we have a 3-paths interferometer and 2 independent phase errors $\delta_1:= \delta \varphi_{0,1}$ and $\delta_2:= \delta \varphi_{0,2}$. The controlled-gates $C(Z_3)$ and $C(Z_3^{err})$ are $9\times 9$ diagonal matrices ($\omega= e^{2\pi i/3}$, $\omega^3=1$, $1+ \omega+ \omega^2= 0$)
\begin{eqnarray}
C(Z_3) &=& \mathrm{diag} (1, 1, 1, 1, \omega, \omega^2, 1, \omega^2, \omega) \\
C(Z_3^{err}) &=& \mathrm{diag} (1, e^{i\delta_1}, e^{i\delta_2}, 1, \omega e^{i\delta_1 \frac{m_1}{m_0}}, \omega^2 e^{i\delta_2 \frac{m_1}{m_0}}, 1, \omega^2 e^{i\delta_1 \frac{m_2}{m_0}}, \omega e^{i\delta_2 \frac{m_2}{m_0}})
\end{eqnarray}

The transition matrix of the imperfect sorter $C(X_3^{err})$ is block-diagonal
\be
C(X_3^{err}) = \mathrm{diag} (E_1, E_2, E_3)
\ee
where the $3\times 3$ matrices $E_k= F^\dag D_k F$ contain the leakage between channels; $F$ is the 3-dimensional Fourier transform and the $D_k$ are the diagonal matrices containing the phase errors in $C(Z_3^{err}) = \mathrm{diag} (D_1, D_2, D_3)$:
\begin{eqnarray}
F &=& \frac{1}{\sqrt 3}\begin{bmatrix} 1 & 1 & 1 \\ 1 & \omega & \omega^2 \\ 1 & \omega^2 & \omega \end{bmatrix} \\
D_1 &=& \mathrm{diag} (1, e^{i\delta_1}, e^{i\delta_2}) \\ 
D_2 &=& \mathrm{diag} (1, \omega\, e^{i\delta_1 \frac{m_1}{m_0}}, \omega^2\, e^{i\delta_2 \frac{m_1}{m_0}}) \\ 
D_3 &=& \mathrm{diag} (1, \omega^2\, e^{i\delta_1 \frac{m_2}{m_0}}, \omega\, e^{i\delta_2 \frac{m_2}{m_0}})
\end{eqnarray}
Notice that $C(X_3^{err})$ is block-diagonal. This means that mass sectors do not mix, i.e., there are no transitions between different masses $m_k \not \rightarrow m_j$. Only the output paths mix: in general a species $m_k$ will be in a superposition of output paths, see eq.~\eqref{CX_error}.

\begin{figure}[t]
  \includegraphics[width= 8cm]{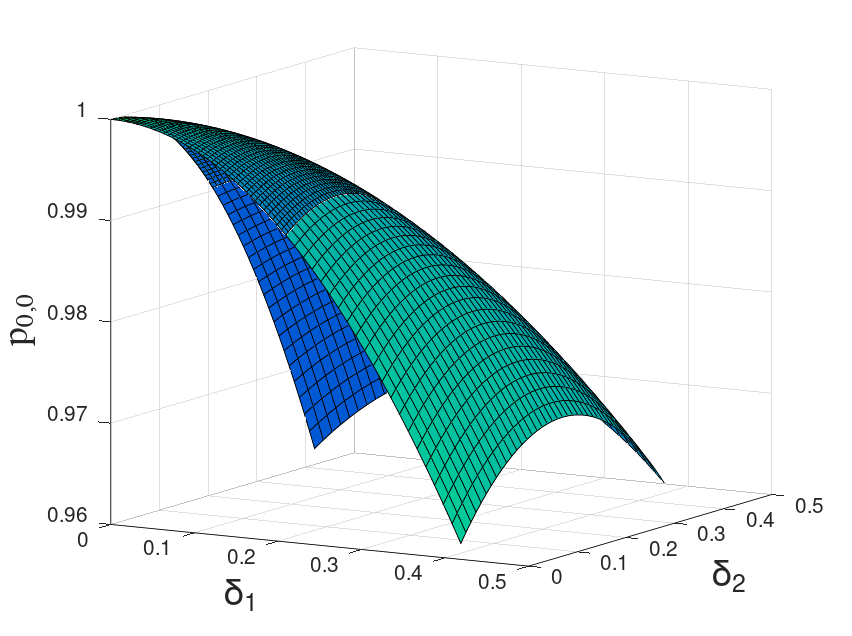}
  \caption{Probability $p_{0,0}$ for species $m_0$ to exit on path 0 as a function of errors $\delta_1, \delta_2$. The probability is $p >0.96$ even for errors as high as $2\pi/15$, representing 20\% of the relevant phase $\delta \varphi= 2\pi/N= 2\pi/3$ for 3 species.}
  \label{p00}
\end{figure}

With the above notations the amplitudes $c_{k, s}$ can be computed analytically:
\begin{eqnarray}
c_{0, 0} = \tfrac{1}{3}(1+ e^{i\delta_1}+ e^{i\delta_2}), \ \ c_{0, 1} = \tfrac{1}{3}(1+ \omega^2 e^{i\delta_1}+ \omega e^{i\delta_2}), \ \ c_{0, 2} = \tfrac{1}{3}(1+ \omega e^{i\delta_1}+ \omega^2 e^{i\delta_2}) \\
c_{1, 0} = \tfrac{1}{3}(1+ \omega e^{i\delta'_1}+ \omega^2 e^{i\delta'_2}), \ \ c_{1, 1} = \tfrac{1}{3}(1+ e^{i\delta'_1}+ e^{i\delta'_2}), \ \ c_{1, 2} = \tfrac{1}{3}(1+ \omega^2 e^{i\delta'_1}+ \omega e^{i\delta'_2}) \\
c_{2, 0} = \tfrac{1}{3}(1+ \omega^2 e^{i\delta''_1}+ \omega e^{i\delta''_2}), \ \ c_{2, 1} = \tfrac{1}{3}(1+ \omega e^{i\delta''_1}+ \omega^2 e^{i\delta''_2}), \ \ c_{2, 2} = \tfrac{1}{3}(1+ e^{i\delta''_1}+ e^{i\delta''_2})
\end{eqnarray}
where $\delta'_k := \delta_k \frac{m_1}{m_0}$, $\delta''_k := \delta_k \frac{m_2}{m_0}$. For $m_0$, the probability of exiting on path $s$ is $p_{0, s}= |c_{0, s}|^2$:
\begin{eqnarray}
p_{0, 0} &=& \tfrac{1}{3}+ \tfrac{2}{9} \left[ \cos \delta_1 + \cos\delta_2+ \cos(\delta_1-\delta_2) \right] \\
p_{0, 1} &=& \tfrac{1}{3}+ \tfrac{2}{9} \left[ \cos (\delta_1- \tfrac{2\pi}{3}) + \cos (\delta_2+ \tfrac{2\pi}{3})+ \cos(\delta_1- \delta_2+ \tfrac{2\pi}{3}) \right] \\
p_{0, 2} &=& \tfrac{1}{3}+ \tfrac{2}{9} \left[ \cos (\delta_1+ \tfrac{2\pi}{3}) + \cos (\delta_2- \tfrac{2\pi}{3})+ \cos(\delta_1- \delta_2- \tfrac{2\pi}{3}) \right]
\end{eqnarray}
In each mass sector the probability is conserved, $\sum_{s=0}^2 p_{k, s}= 1, \forall k$. We obtain similar formulae for the output probabilities for $m_1, m_2$. In the ideal case, $\delta_1= \delta_2= 0$ and $p_{0, 0}= 1$, $p_{0, 1}= p_{0, 2}= 0$, as expected.

For $m_0$ the leakage to other channels is $1- p_{0,0}$. As we can see from Fig.~\ref{p00}, the leakage is less than 4\% even for phase errors $\delta_{1,2}= \frac{2\pi}{15}$, which are 20\% of the relevant phase for 3 species $\delta \varphi= \frac{2\pi}{N}=\frac{2\pi}{3}$. Furthermore, for phase errors  $\le \frac{\delta \varphi}{10}$, the leakage is less than 1\%.

\section{Discussion and conclusions}

It is insightful to compare interferometric mass spectrometry proposed here with the standard, widely-used accelerator mass spectrometry.

Conceptually, the two methods are complementary: in AMS samples are particle-like with localised, well-defined trajectories, while in Interf-MS they have a wave-like behaviour with delocalised trajectories. More importantly, this duality also extends to the velocity regime. In AMS the samples are accelerated to high velocities. In contrast, Interf-MS uses the low-velocity regime, corresponding to large $\lambda$. The differences between the two methods are summarised in Table \ref{compare}.

\noindent{\bf Separation mechanism.} A crucial difference between AMS and Interf-MS is the separation mechanism. AMS uses the Lorentz force $q(\bf E + v \times B)$ for separation. This has two implications: (i) the samples have to be electrically charged, and (ii) we need electric and magnetic fields. Hence AMS does not work for neutral particles (atoms, molecules). In contrast, since Interf-MS uses quantum interference for sorting, it works also for neutral samples. More importantly, since Interf-MS does not use electromagnetic fields, it dispenses with bulky magnets.

\noindent{\bf Measurement.} AMS measures only the mass-to-charge ratio $m/q$ since it uses the Lorentz force. In contrast, Interf-MS measures directly the mass $m$.

\noindent{\bf Velocity regime.} In AMS the separation between mass species is proportional to the velocity. In order to achieve a very large dynamical range, AMS uses accelerators (e.g., TANDEMs). In contrast, Interf-MS works in the low-velocity regime, so accelerators are not necessary. Consequently, Interf-MS can lead to compact devices, with low size and weight. This opens the possibility of portable mass spectrometers for field work.

\noindent{\bf Velocity filters.} Both methods use velocity filters. Since Interf-MS works in the low-velocity regime, one can use mechanical choppers. On the other hand, mechanical choppers cannot be used in the high-velocity regime of AMS, as they would introduce large errors.

\begin{table}[h]
\begin{tabular}{p{4.2cm}|p{3cm} p{3cm}}
&\ \  AMS & Interf-MS \\
\hline
quantum behaviour & particle & wave \\
trajectory & localised & delocalised \\
separation mechanism & Lorentz force & interference \\
measurement & $m/q$ & $m$ \\
separation & $\sim v$ & $\sim v^{-1}$ \\
velocity regime & large $v$ & small $v$ \\
\end{tabular}
\caption{Complementarity between accelerator mass spectrometry (AMS) and interferometric mass spectrometry (Interf-MS).}
\label{compare}
\end{table}

In conclusion, due to this complementarity Interf-MS can potentially be used for applications where AMS is impractical. These include sensitive samples (e.g., biological, proteins) that break upon acceleration (or at the stripping phase in TANDEMs) and neutral samples which are difficult to ionise. Moreover, since accelerators are not required, we envisage that Interf-MS will lead to compact devices for mobile applications.

\begin{acknowledgement}

The author thanks Rare\c s \c Suv\u ail\u a for many stimulating discussions. This work has been supported from a grant of the Romanian Ministry of Research and Innovation, PCCDI-UEFISCDI, Project Number PN-III-P1-1.2-PCCDI-2017-0338/79PCCDI/2018, within PNCDI III, and from PN 19060101/2019-2022.

\end{acknowledgement}

%%%%%%%%%%%%%%%%%%%%%%%%%%%%%%%%%%%%%%%%%%%%

\end{document}